\documentclass{PoS}

\title{Production of charm quark/antiquark pairs at LHC}

\ShortTitle{Production of $c \bar c$ pairs at LHC}

\author{\speaker{Rafal Maciula}
\thanks{This work was supported in part by the Polish MNiSW grants No. N N202 237040 and DEC-2011/01/B/ST2/04535.}\\
        Institute of Nuclear
Physics PAN, PL-31-342 Cracow, Poland\\
        E-mail: \email{rafal.maciula@ifj.edu.pl}}

\author{Marta Luszczak\\
        University of Rzesz\'ow, PL-35-959 Rzesz\'ow, Poland\\
        E-mail: \email{luszczak@univ.rzeszow.pl}}
        
\author{Antoni Szczurek\\
        Institute of Nuclear
Physics PAN, PL-31-342 Cracow, Poland\\ \& University of Rzesz\'ow, PL-35-959 Rzesz\'ow, Poland\\
        E-mail: \email{antoni.szczurek@ifj.edu.pl}}

\abstract{We discuss charm production at LHC. The production of single $c \bar c$ pairs
is calculated in the $k_t$-factorization approach. We use several
unintegrated gluon distributions from the literature. The hadronization is included with the help of
fragmentation functions found for the production of charmed mesons in $e^+ e^-$ collisions. Differential distributions for several 
charmed mesons are presented and compared to recent results of the ALICE
and LHCb collaborations. Furthermore we discuss production of two pairs of $c \bar c$ within a simple
formalism of double-parton scattering (DPS). Surprisingly large
cross sections, comparable to single-parton scattering (SPS)
contribution, are predicted for LHC energies.
Both total inclusive cross section as a function of energy and
differential distributions are shown.
We include recently discussed evolution of double partons in the case
of two scales. We discuss perspectives how to identify the double scattering
contribution. We find much larger cross section for large rapidity distance
between charm quarks from different hard parton scatterings
compared to single scattering.}

\FullConference{Sixth International Conference on Quarks and Nuclear Physics,\\
		April 16-20, 2012\\
		Ecole Polytechnique, Palaiseau, Paris}

\begin{document}

\section{Transverse momentum spectra of open charm mesons at LHC}

Recently ALICE and LHCb collaborations have measured inclusive transverse momentum spectra
of open charm mesons in proton-proton collisions 
at $\sqrt{s}=7$ TeV \cite{ALICE,LHCb}. These measurements are very interesting from the theoretical point of view because of the collision energy never achieved before and unique rapidity acceptance of the detectors. Especially, results from forward rapidity region $2 < y < 4$, obtained by LHCb can improved our understanding of pQCD production of heavy quarks.

The inclusive production of heavy quark/antiquark pairs can be calculated
in the framework of the $k_t$-factorization \cite{CCH91}.
In this approach transverse momenta of initial partons are included and
emission of gluons is encoded in a so-called unintegrated gluon,
in general parton, distributions (UGDFs).
In the leading-order approximation (LO) within the $k_t$-factorization approach
the differential cross section for the $Q \bar Q$
can be written as:
\begin{eqnarray}
\frac{d \sigma}{d y_1 d p_{1t} d y_{2} d p_{2t} d \phi} =
\sum_{i,j} \; \int \frac{d^2 \kappa_{1,t}}{\pi} \frac{d^2 \kappa_{2,t}}{\pi}
\frac{1}{16 \pi^2 (x_1 x_2 s)^2} \; \overline{ | {\cal M}_{ij} |^2}\\
\nonumber 
\delta^{2} \left( \vec{\kappa}_{1,t} + \vec{\kappa}_{2,t} 
                 - \vec{p}_{1,t} - \vec{p}_{2,t} \right) \;
{\cal F}_i(x_1,\kappa_{1,t}^2) \; {\cal F}_j(x_2,\kappa_{2,t}^2) \; , \nonumber \,\,
\end{eqnarray}
where ${\cal F}_i(x_1,\kappa_{1,t}^2)$ and ${\cal F}_j(x_2,\kappa_{2,t}^2)$
are the unintegrated gluon (parton) distribution functions. 

There are two types of the LO $2 \to 2$ subprocesses which contribute
to heavy quarks production, $gg \to Q \bar Q$ and $q \bar q \to Q \bar
Q$. The first mechanism dominates at large energies and the second one
near the threshold. Only $g g \to Q \bar Q$ mechanism is included here. We use off-shell matrix elements
corresponding to off-shell kinematics so hard amplitude depends on transverse momenta
(virtualities of initial gluons). At very high energies, especially in the case of charm production at forward rapidities, rather small $x$-values become relevant. In this kinematical regime calculation of unintegrated parton distributions is not under full theoretical control and can include in different ways various theoretical aspects, like effect of small-$x$ saturation or treatment of nonperturbative region. In order to show the uncertainty of our predictions resulting from
different approaches in calculating unintegrated parton distributions we have used several models from the literature. All of them have different theoretical background.
It is therefore very interesting to compare such results with the recent ALICE and LHCb data 
and verify applicability of these UGDFs at LHC energies. More details of theoretical model applied here can be found in Ref.~\cite{LMS09}.

The hadronization of heavy quarks is usually done
with the help of fragmentation functions. The inclusive distributions of
hadrons can be obtained through a convolution of inclusive distributions
of heavy quarks/antiquarks and Q $\to$ h fragmentation functions:
\begin{equation}
\frac{d \sigma(y_h,p_{t,h})}{d y_h d^2 p_{t,h}} \approx
\int_0^1 \frac{dz}{z^2} D_{Q \to h}(z)
\frac{d \sigma_{g g \to Q}^{A}(y_Q,p_{t,Q})}{d y_Q d^2 p_{t,Q}}
\Bigg\vert_{y_Q = y_h \atop p_{t,Q} = p_{t,h}/z}
 \; ,
\label{Q_to_h}
\end{equation}
where $p_{t,Q} = \frac{p_{t,h}}{z}$, where $z$ is the fraction of longitudinal momentum of heavy quark carried by meson.
We have made approximation assuming that $y_{Q}$  is
unchanged in the fragmentation process.

\begin{figure}[!h]
\begin{minipage}{0.47\textwidth}
 \centerline{\includegraphics[width=1.0\textwidth]{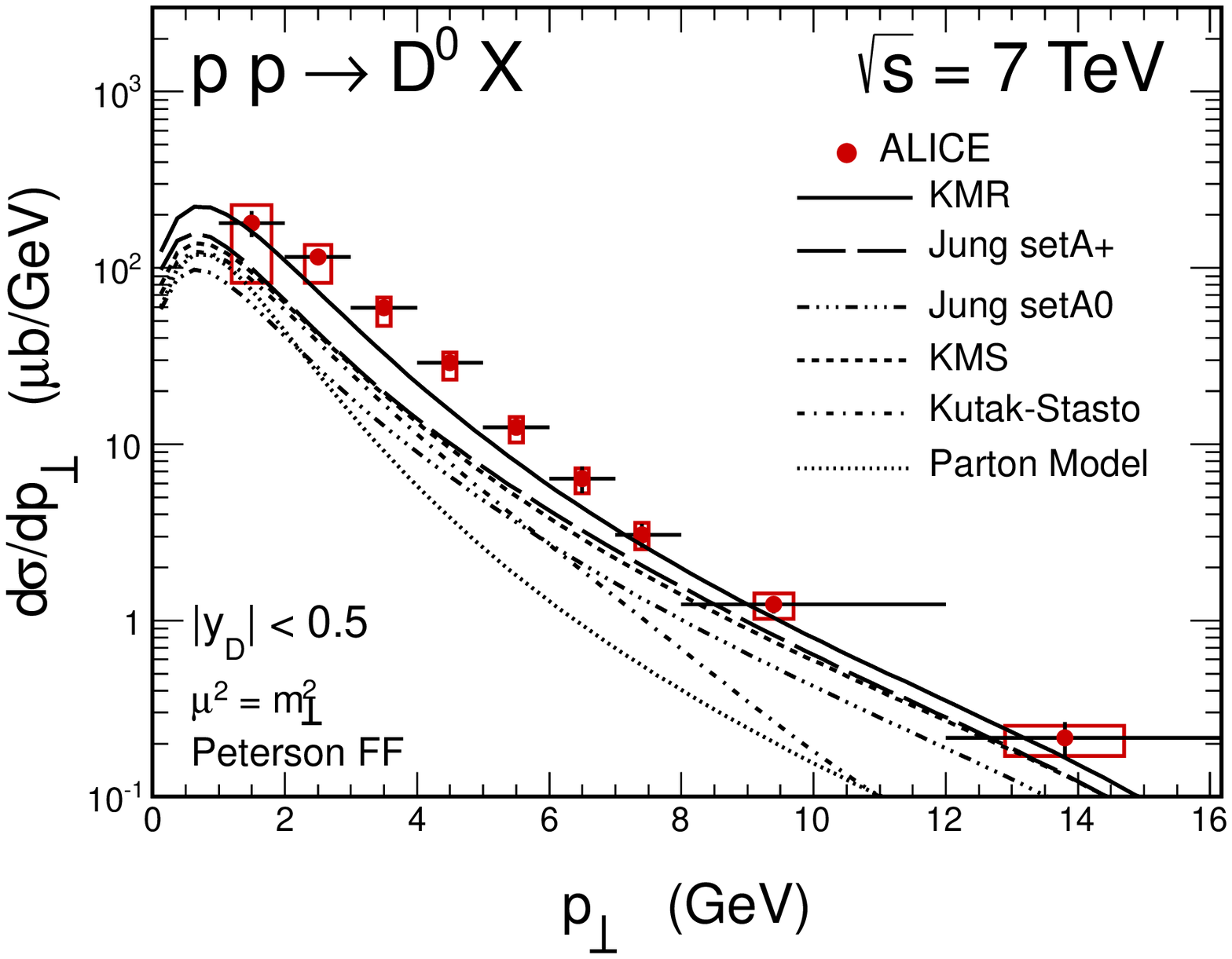}}
\end{minipage}
\hspace{0.5cm}
\begin{minipage}{0.47\textwidth}
 \centerline{\includegraphics[width=1.0\textwidth]{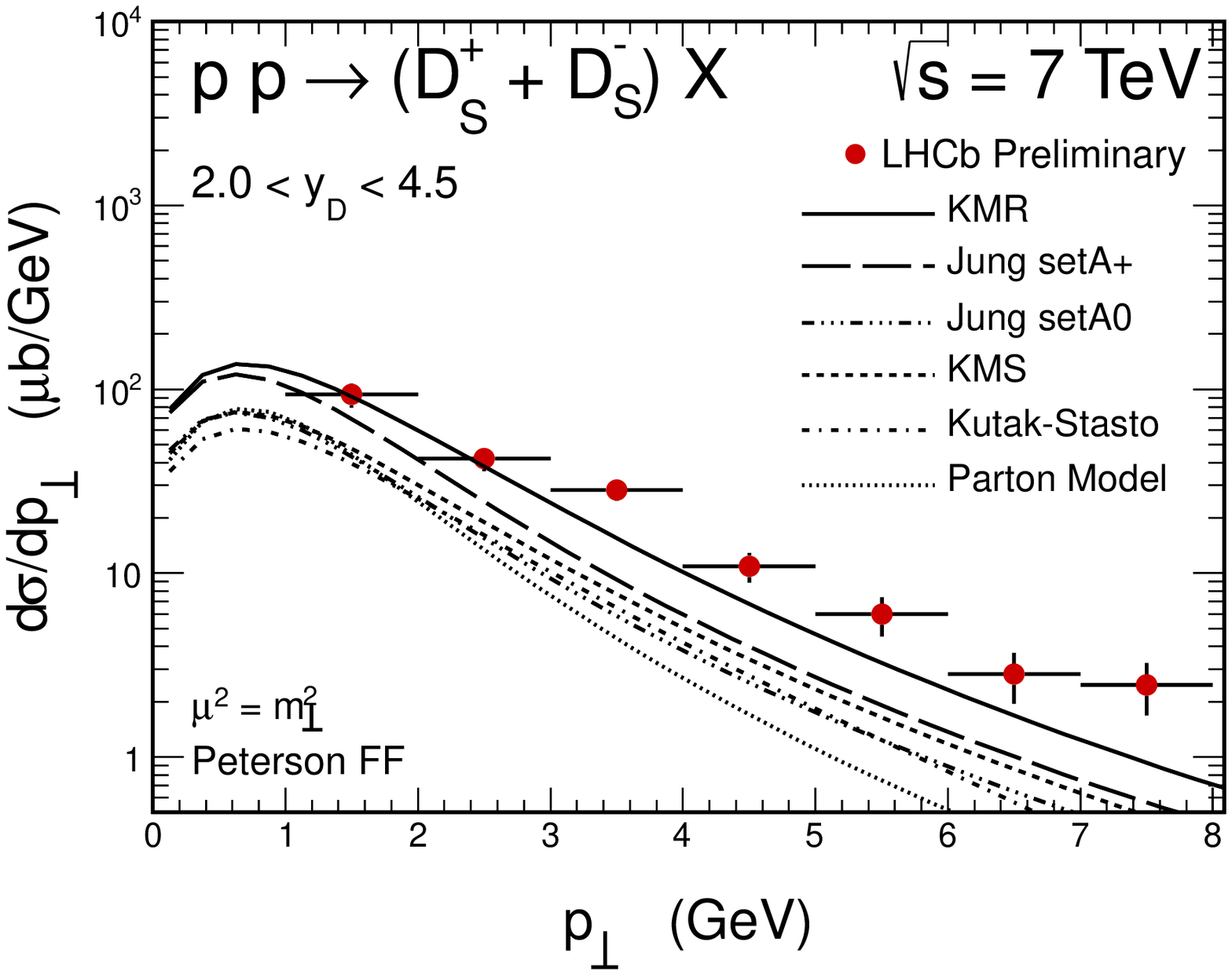}}
\end{minipage}
   \caption{
\small Transverse momentum distributions of D mesons together with the ALICE (left) and LHCb (right) data. Each curve corresponds to different model of UGDFs. The calculations were done with the help of Peterson fragmentation function.}
 \label{fig:pt-alice-D-1}
\end{figure}

In Fig.~\ref{fig:pt-alice-D-1} we present our predictions for differential distributions in transverse momentum of open charm mesons together with
the ALICE (left panel) and LHCb (right panel) experimental data. We plot here results obtained with different models of UGDFs.
The calculations of charm quarks are performed for $m_c = 1.5$ GeV with the values of the
renormalization and factorization scales taken to be $\mu_{R}^2=\mu_{F}^2=m_{t}^2$. We get very good description of the experimental data, in both ALICE and LHCb cases only with KMR model of unintegrated gluon distributions. The other of the applied parametrizations of UGDFs do not work and clearly underestimate experimental data points.
One can also observe huge difference between results of standard collinear LO parton model predictions (dotted line) and those obtained in the LO $k_t$-factorization approach (solid line).

\begin{figure}[!h]
\begin{minipage}{0.47\textwidth}
 \centerline{\includegraphics[width=1.0\textwidth]{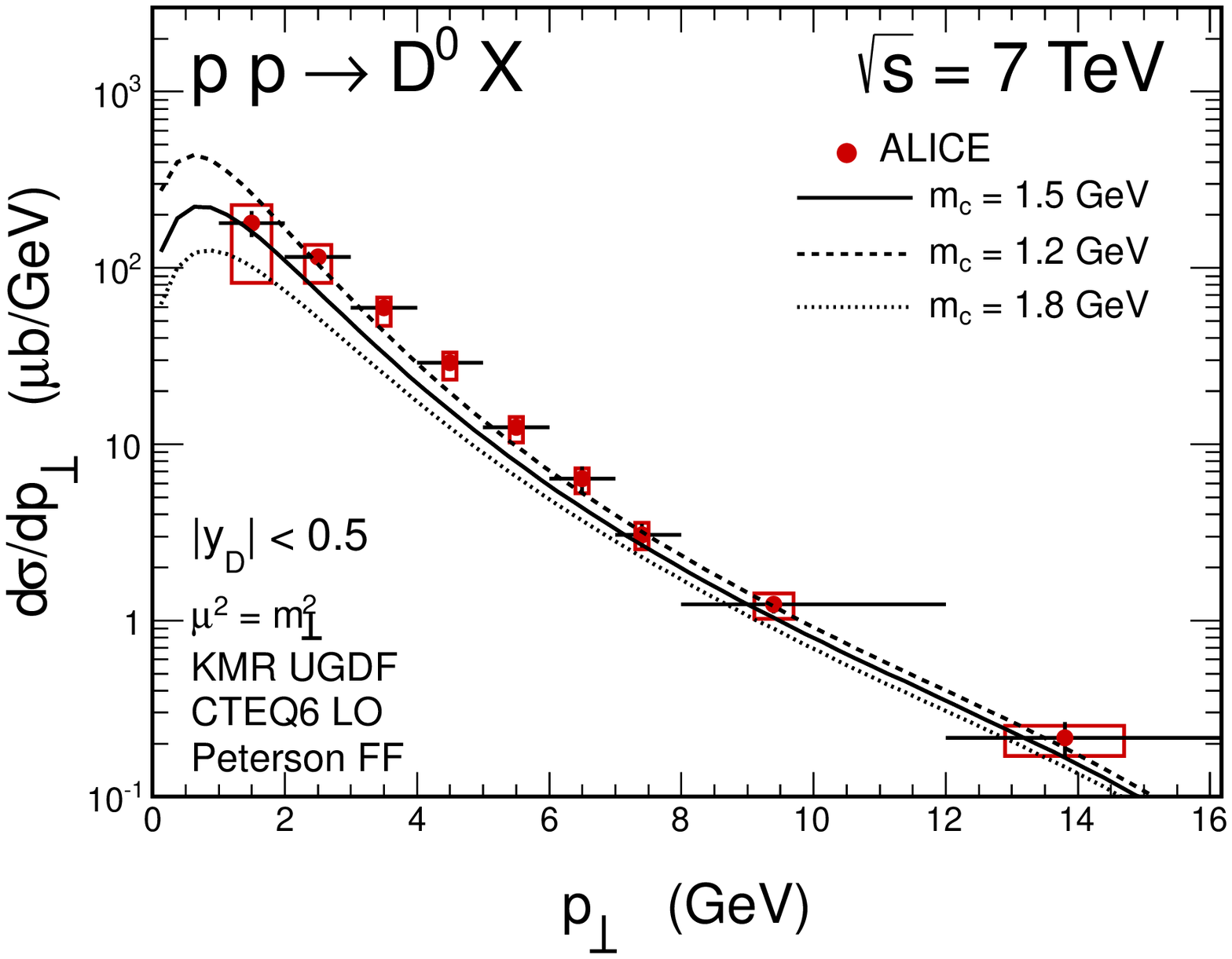}}
\end{minipage}
\hspace{0.5cm}
\begin{minipage}{0.47\textwidth}
 \centerline{\includegraphics[width=1.0\textwidth]{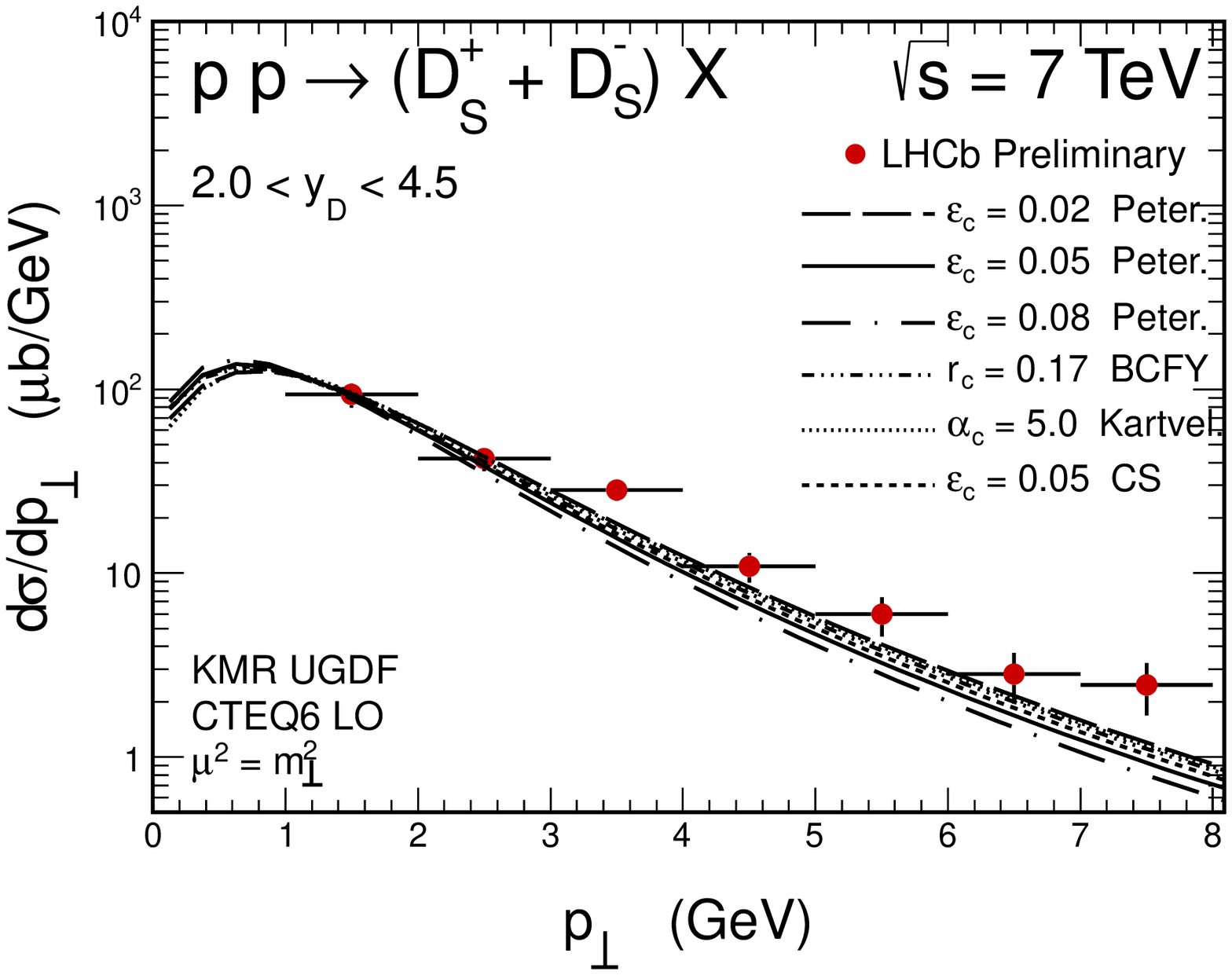}}
\end{minipage}
   \caption{
\small  Uncertainties of our predictions related to the charm quark mass (left) and to the choice of fragmentation function (right).}
 \label{fig:pt-alice-D-2}
\end{figure}

In Fig.\ref{fig:pt-alice-D-2} we discuss uncertainties due to the 
modification of the charm quark mass $m_c \in$ (1.2, 1.8 GeV) (left panel), as well as uncertainties related to different models of heavy quark fragmentation (right panel).
As one can observe, there is only a small sensitivity of the results 
to the value of charm quark mass at small transverse momenta. Some small uncertainties due to the choice of fragmentation function appear only at larger values of meson $p_{t}$'s.

\section{Double charm production via Double Parton Scattering}

\begin{figure}[!h]
\begin{center}
\includegraphics[width=4cm]{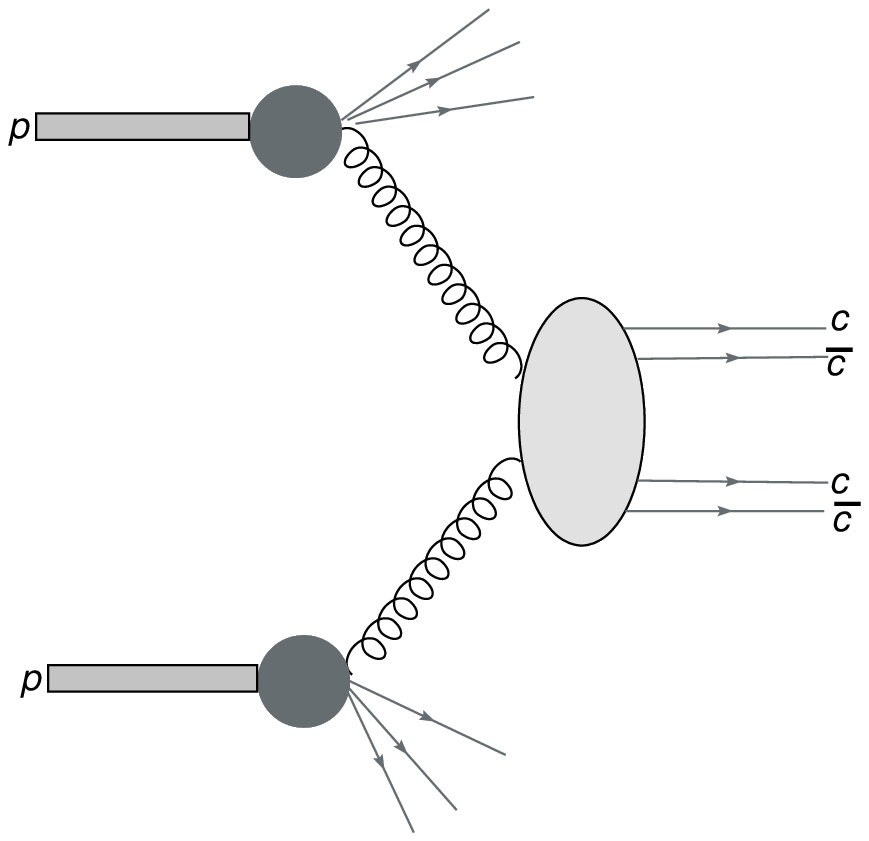}
\includegraphics[width=4cm]{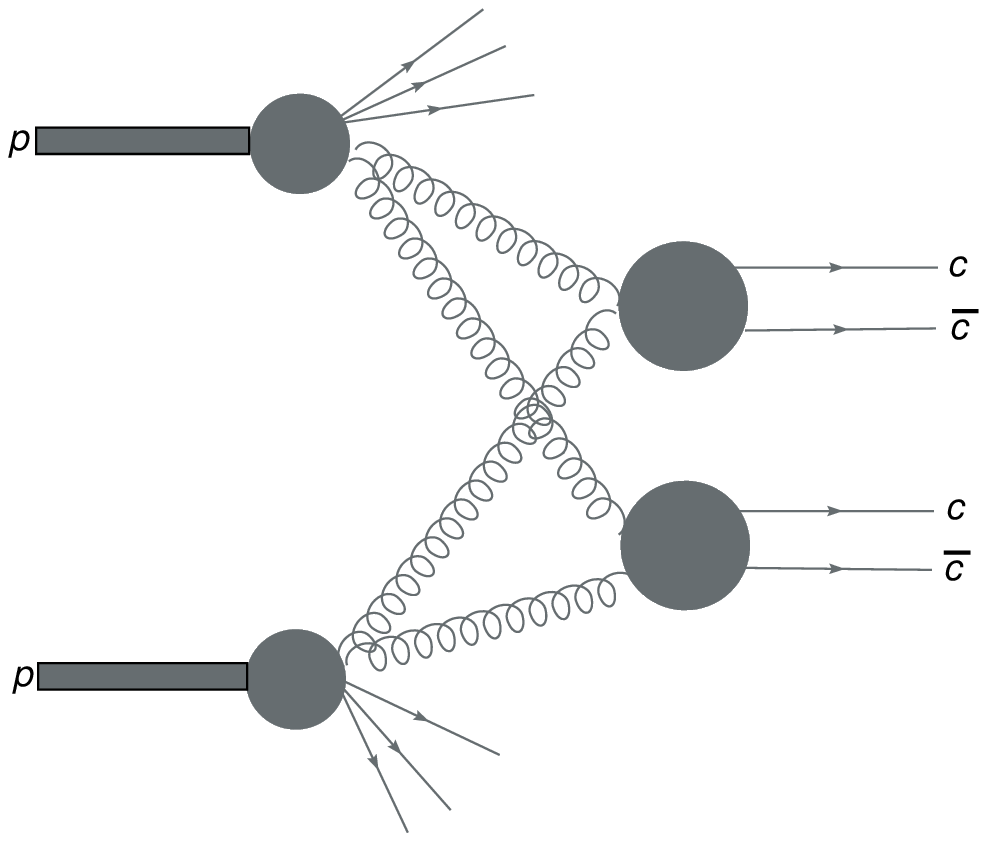}
\end{center}
   \caption{
\small SPS (left) and DPS (right) mechanisms of $(c \bar c) (c \bar c)$ 
production.  
}
 \label{fig:diagram}
\end{figure}

The mechanism of double-parton scattering (DPS) production of two pairs of
heavy quark and heavy antiquark is shown in Fig.~\ref{fig:diagram}
together with corresponding mechanism of single-scattering production.
The double-parton scattering has been recognized and discussed already 
in seventies and eighties. The activity stopped when it was realized that their contribution at center-of-mass energies available then was negligible. Nowadays, the theory of the double-parton
scattering is quickly developing (see e.g. \cite{S2003,KS2004,GS2010})
which is partly driven by new results from the LHC.

In the present analysis we wish to concentrate on the production of $(c \bar c)
(c \bar c)$ four-parton final state which has not been carefully discussed so
far, but, as will be shown here, is particularly interesting especially in 
the context of experiments being carried out at LHC and/or high-energy atmospheric
and cosmogenic neutrinos (antineutrinos).

The double-parton scattering formalism in the simplest form assumes two
single-parton scatterings. Then in a simple probabilistic picture the
cross section for double-parton scattering can be written as:
\begin{equation}
\sigma^{DPS}(p p \to c \bar c c \bar c X) = \frac{1}{2 \sigma_{eff}}
\sigma^{SPS}(p p \to c \bar c X_1) \cdot \sigma^{SPS}(p p \to c \bar c X_2).
\label{basic_formula}
\end{equation}
This formula assumes that the two subprocesses are not correlated and do
not interfere.
At low energies one has to include parton momentum conservation
i.e. extra limitations: $x_1+x_3 <$ 1 and $x_2+x_4 <$ 1, where $x_1$ and $x_3$
are longitudinal momentum fractions of gluons emitted from one proton and $x_2$ and $x_4$
their counterparts for gluons emitted from the second proton. The "second"
emission must take into account that some momentum was used up in the "first" parton
collision. This effect is important at large quark or antiquark rapidities.
Experimental data \cite{Tevatron} provide an estimate of $\sigma_{eff}$
in the denominator of formula (\ref{basic_formula}). In our analysis we
take $\sigma_{eff}$ = 15 mb.

A more general formula for the cross section can be written formally 
in terms of double-parton distributions (dPDF), e.g. $F_{gg}$, $F_{qq}$, etc. 
In the case of heavy quark production at high energies:
\begin{eqnarray}
d \sigma^{DPS} &=& \frac{1}{2 \sigma_{eff}}
F_{gg}(x_1,x_3,\mu_1^2,\mu_2^2) F_{gg}(x_2,x_4,\mu_1^2,\mu_2^2)
\nonumber \\
&&d \sigma_{gg \to c \bar c}(x_1,x_2,\mu_1^2)
d \sigma_{gg \to c \bar c}(x_3,x_4,\mu_2^2) \; dx_1 dx_2 dx_3 dx_4 \, .
\label{cs_via_doublePDFs}
\end{eqnarray}
It is physically motivated to write the double-parton distributions
rather in the impact parameter space 
$F_{gg}(x_1,x_2,b) = g(x_1) g(x_2) F(b)$, where $g$ are 
usual conventional parton distributions and $F(b)$ is an overlap of 
the matter distribution in the 
transverse plane where $b$ is a distance between both gluons
\cite{CT1999}. The effective cross section in 
(\ref{basic_formula}) is then $1/\sigma_{eff} = \int d^2b F^2(b)$ and 
in this approximation is energy independent.

The double-parton distributions in Eq.(\ref{cs_via_doublePDFs})
are generally unknown. Usually one assumes a factorized form and
expresses them via standard distributions for SPS.
Even if factorization is valid at some scale, QCD evolution may lead
to a factorization breaking. For some time the evolution was known only 
when the scale of both scatterings is the same \cite{GS2010}
i.e. for heavy object, like double gauge boson production.
Recently the evolution of dPDF was discussed also in the case of 
different scales \cite{C11}. 

\begin{figure}[!h]
\begin{minipage}{0.47\textwidth}
 \centerline{\includegraphics[width=1.0\textwidth]{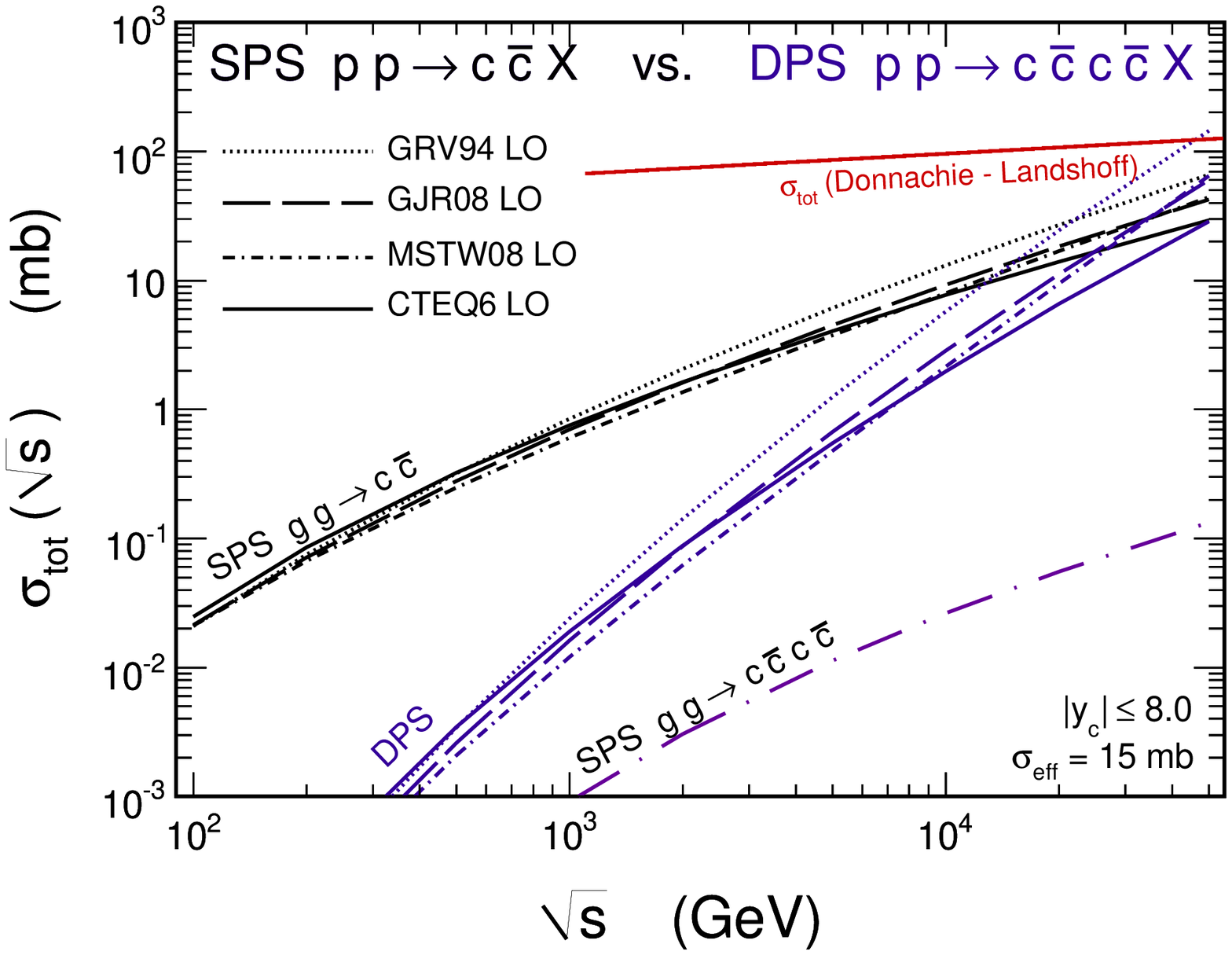}}
\end{minipage}
\hspace{0.5cm}
\begin{minipage}{0.47\textwidth}
 \centerline{\includegraphics[width=1.0\textwidth]{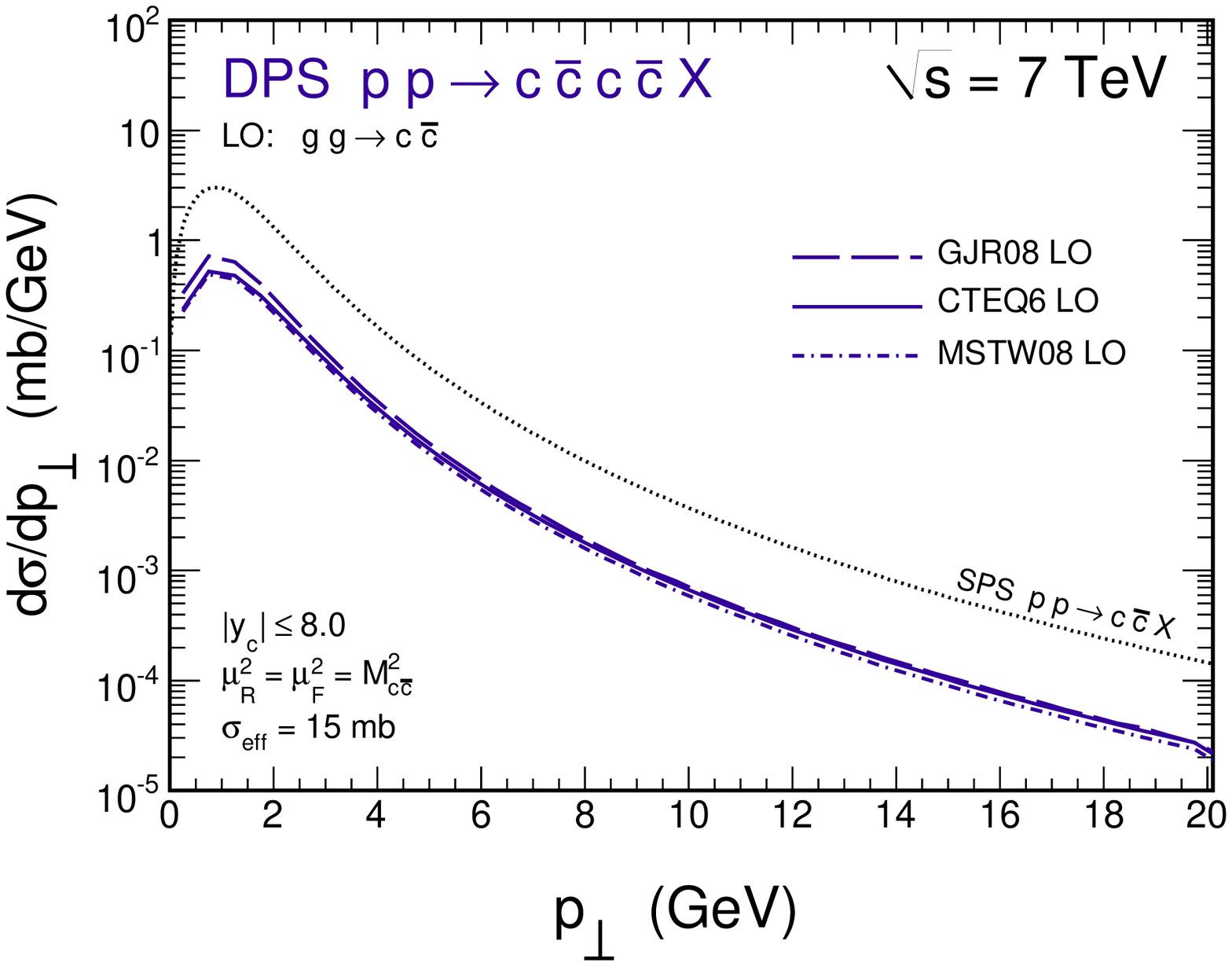}}
\end{minipage}
   \caption{
\small Total LO cross section for single
$c \bar c$ pair and SPS and DPS $c \bar c c \bar c$ production as a function of center-of-mass energy (left panel) and  
transverse momentum (right panel) of $c$ or $\bar{c}$ quarks  at $\sqrt{s}$ = 7 TeV.
Cross section for DPS should be multiplied in addition by a factor 2 in the case when all $c$ ($\bar c$) are counted.
We show in addition a parametrization of the total cross section in the left panel.
}
 \label{fig:single_vs_double_LO}
\end{figure}

In the left panel of Fig.~\ref{fig:single_vs_double_LO} we
compare cross sections for the single $c \bar c$ pair production as well
as for single-parton and double-parton scattering $c \bar c c \bar c$
production as a function of proton-proton center-of-mass energy. 
At low energies the conventional single $c \bar c$ pair production
dominates. The cross section for SPS production
of $c \bar c c \bar c$ system \cite{SS2012} is more than two orders of magnitude smaller
than that for single $c \bar c$ production. For reference we show the
proton-proton total cross section as a function of energy as
parametrizes in Ref.~\cite{DL92}. At low energy the $c \bar c$ or $ c \bar c c \bar c$ cross sections are much
smaller than the total cross section. At higher energies the contributions
approach the total cross section. This shows that inclusion of
unitarity effect and/or saturation of parton distributions may be necessary.
At LHC energies the cross section for both terms becomes comparable.
This is a new situation when the DPS gives a huge contribution to inclusive
charm production. 

In the right panel of Fig.~\ref{fig:single_vs_double_LO} we present
transverse momentum distributions of single $c$ ($\bar c$). Within approximations made in this
paper the distributions are identical in shape to single-pair production
distributions. This means that the double-scattering contribution
produces naturally an extra center-of-mass energy dependent $K$ factor
to be contrasted with approximately energy-independent $K$-factor due to
next-to-leading order QCD corrections. Other interesting conclusions
can be obtained by studying correlation observables.

\begin{figure}[!h]
\begin{minipage}{0.47\textwidth}
 \centerline{\includegraphics[width=1.0\textwidth]{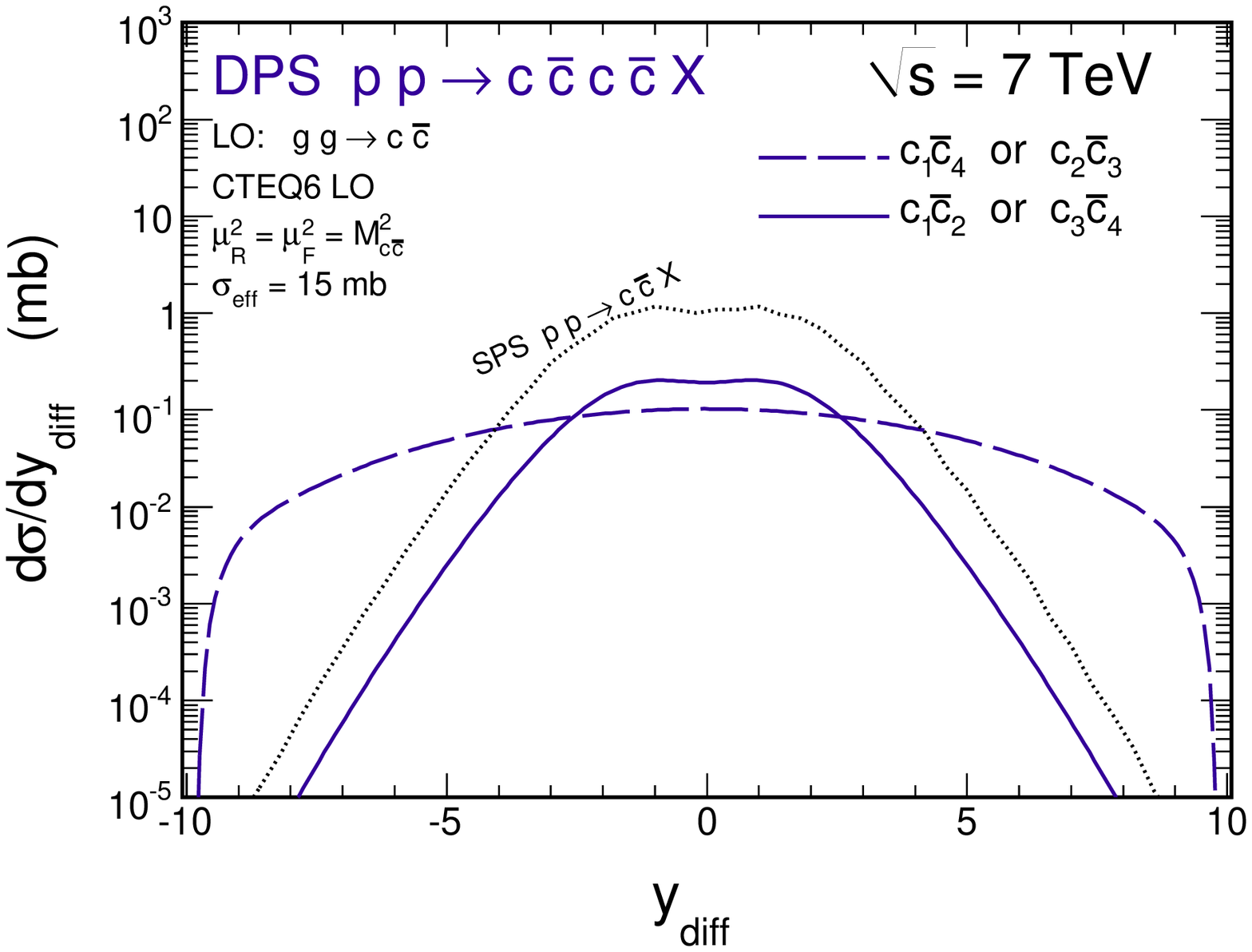}}
\end{minipage}
\hspace{0.5cm}
\begin{minipage}{0.47\textwidth}
 \centerline{\includegraphics[width=1.0\textwidth]{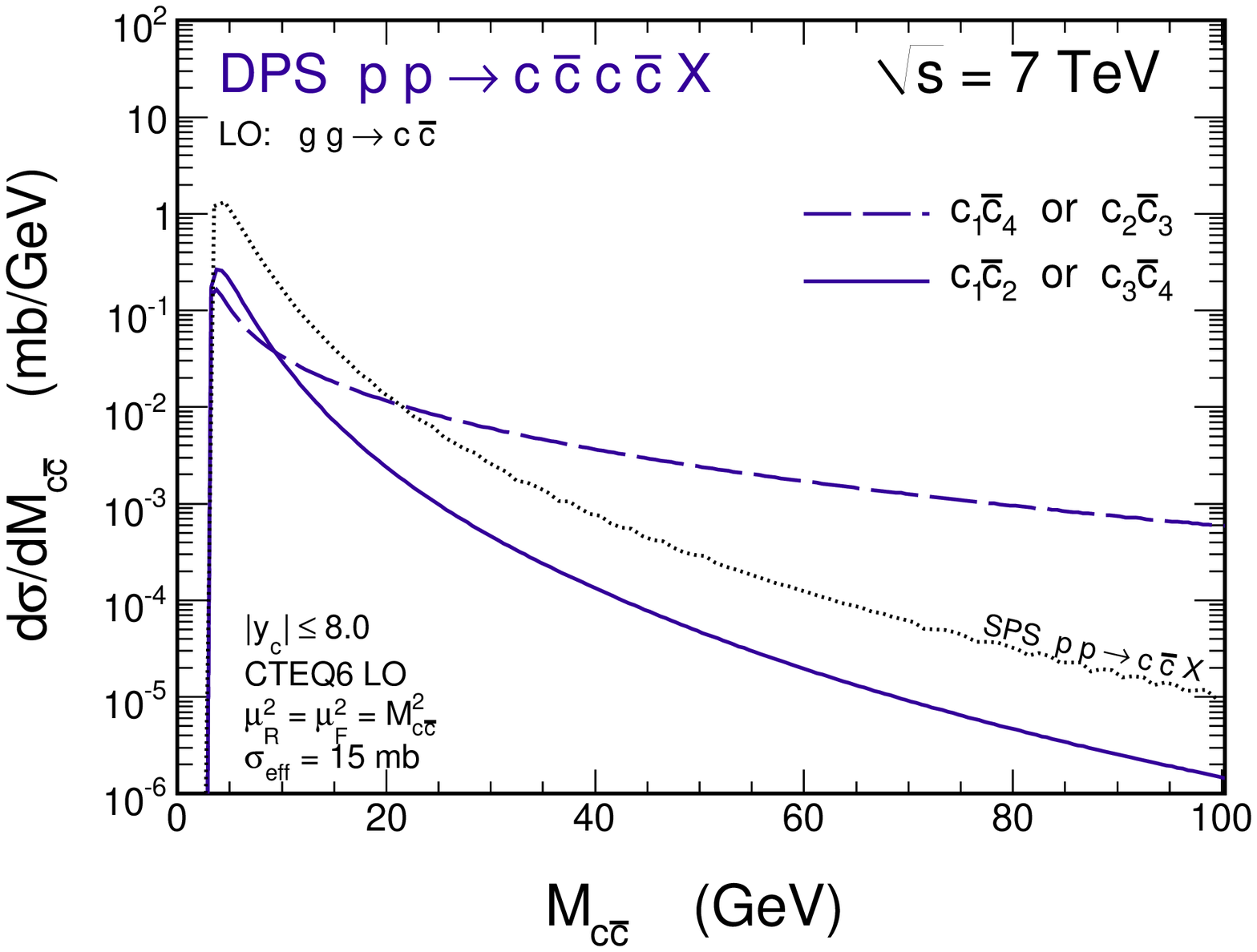}}
\end{minipage}
   \caption{
\small Distribution in rapidity difference (left panel) and in invariant
mass of the $c\bar{c}$ pair (right panel) at $\sqrt{s}$ = 7 TeV.
}
\label{fig:double_correlations_1}
\end{figure}

In Fig.~\ref{fig:double_correlations_1} we show distribution in the
difference of $c$ and $\bar c$ rapidities $y_{diff} = y_c - y_{\bar c}$
(left panel) as well as in the $c \bar c$
invariant mass $M_{c\bar c}$ (right panel). We show both terms: when 
$c \bar c$ are emitted in the same parton scattering 
($c_1\bar c_2$ or $c_3\bar c_4$) and when they are emitted from different 
parton scatterings ($c_1\bar c_4$ or $c_2\bar c_3$). In the latter case
we observe a long tail for large rapidity difference as well as at large
invariant masses of $c \bar c$.

In particular, $c c$ (or $\bar c \bar c$) should be predominantly
produced from two different parton scatterings which opens a possibility 
to study the double scattering processes.
A good signature of the $c \bar c c \bar c$ final state is a production
of two mesons, both containing $c$ quark or two mesons
both containing $\bar c$ antiquark ($D^0 D^0$ or/and ${\bar D}^0 {\bar D}^0$) in one physical event.
More detailed discussion of the DPS charm production can be found in our original paper Ref.~\cite{LMS2012}.

In the present approach we have calculated cross section in a simple
collinear leading-order approach. A better approximation would be to include
multiple gluon emissions. This can be done e.g. in soft gluon resummation
or in $k_t$-factorization approach. This will be
discussed in detail elsewhere.


\begin{thebibliography}{99}

\bibitem{ALICE}
B. Abelev et al. (The ALICE Collaboration), J. High Energy Phys. {\bf 01} (2012) 128.

\bibitem{LHCb}
The LHCb Collaboration, LHCb-CONF-2010-013.

\bibitem{CCH91}
S. Catani, M. Ciafaloni and F. Hautmann, Nucl. Phys. {\bf 366} (1991) 135.

\bibitem{LMS09}
M. {\L}uszczak, R. Maciu{\l}a and A. Szczurek, Phys. Rev. {\bf D79} (2009) 034009.

\bibitem{S2003}
A.M. Snigirev, Phys. Rev. {\bf D68} (2003) 114012.

\bibitem{KS2004}
V.L. Korotkikh and A.M. Snigirev, Phys. Lett. {\bf B594} (2004) 171.

\bibitem{GS2010}
J.R. Gaunt and W.J. Stirling, J. High Energy Phys. {\bf 03} (2010) 005.

\bibitem{Tevatron}
F. Abe et al. (The CDF Collaboration), Phys. Rev. {\bf D56} (1997) 3811.

\bibitem{CT1999}
G. Calucci and D. Treleani, Phys. Rev. {\bf D60} (1999) 054023.

\bibitem{C11}
F.A. Ceccopieri, Phys. Lett. {\bf B697} (2011) 482.

\bibitem{SS2012}
W. Sch\"afer and A. Szczurek, Phys. Rev. {\bf D85} (2012) 094029.

\bibitem{DL92}
A. Donnachie and P.V. Landshoff, Phys. Lett. {\bf B296} (1992) 227.

\bibitem{LMS2012}
M. {\L}uszczak, R. Maciu{\l}a and A. Szczurek, Phys. Rev. {\bf D85} (2012) 094034.

\end{thebibliography}
\end{document}